\documentclass[manuscript]{acmart}
\AtBeginDocument{%
  }

\setcopyright{acmlicensed}
\copyrightyear{2026}
\acmYear{2026}
\acmDOI{XXXXXXX.XXXXXXX}

\usepackage{titlesec}
\usepackage{multirow}
\usepackage{array}

\begin{document}

\title{Regimes of Scale in AI Meteorology}


\author{Anya Martin}
\orcid{0000-0002-9592-4024}
\affiliation{%
  \institution{Georgia Institute of Technology}
  \city{Atlanta}
  \country{USA}
}
\email{amartin388@gatech.edu}

\author{Cindy Lin}
\email{clin646@gatech.edu}
\orcid{}
\affiliation{
  \institution{Georgia Institute of Technology}
  \city{Atlanta}
  \state{Georgia}
  \country{USA}
}

\renewcommand{\shortauthors}{Martin et al.}

\begin{abstract}

HCI work has explored the effective integration of AI/ML tools across application domains from healthcare to finance to transportation. We add to this literature with an analysis of AI/ML tools in meteorology, a domain that already uses big data and massive physics-based models. Drawing from 18 interviews with forecasters and meteorologists with varied connections to AI/ML weather modeling, we trace tensions in AI/ML weather application arising from what we call \textbf{regimes of scale}, different ways that AI/ML and meteorological systems make observations, data, and models scale. Rather than seeing AI/ML as a domain-agnostic tool, we argue that AI/ML methods were born from specific platform and internet infrastructures, and so they can struggle to integrate with very different (in this case meteorological) ways of organizing data pipelines.

\end{abstract}



\received{26 Aug 2025}

\maketitle
\section{Introduction}

Human-Computer Interaction (HCI) research has explored the effective integration of artificial intelligence and machine learning (AI/ML) tools across application domains from healthcare \cite{yoo2025} to finance \cite{wang2025} to transportation \cite{yildirim2023} to investigative data journalism and legal analysis \cite{showkat2022}. These domains, alongside their domain knowledge and domain experts, form the application space for AI/ML tools and methods. As HCI scholar David Ribes argues, the term domain defines "spheres of worldly action or knowledge" traversed by highly mobile domain-independent methods \cite{ribes2019}, previously data science and currently AI/ML. HCI studies of AI-in-practice have focused on two major sites of application friction: general AI frameworks and their conflicts with specialized user needs \cite{gurita2025,yoo2025}, and general AI ways of knowing and their conflicts with domain experts' situated knowledge and values \cite{jung2022,jung2024}. We focus on a third site of friction -- differences in \textbf{regimes of scale} between AI/ML and differently-organized Big Data systems -- and provide a case study of AI/ML applications in meteorology, a domain which already has big data and global-scale physics models.

This paper's central framework is \textbf{regimes of scale}. A regime of scale involves the organizational strategy, infrastructural design, and sense of good and/or bad data that allows data to be produced, maintained, and used at scale. Meteorology's regime of scale, which we call \textbf{state scale}, gathers spatially-embedded data to render territory legible across national borders, creating global-scale observations, data, and weather/climate models which have steadily improved over the last seventy years \cite{bauer2015}. This scale is expressed in spatial and temporal terms (mesoscale, synoptic scale, global scale), and smaller scales operate within larger ones via basic physical laws; ideally, this means models can become "big without changing" \cite{tsing2011}. 

In contrast, the \textbf{entrepreneurial scale} of AI/ML is typically expressed in a human dimension: startups aim to capture more users and "scale up" by increasing the amount of data they can collect and analyze from individuals \cite{seaver2021}, while large tech companies utilize their control of key information nodes like Github or Google Search to train large-scale models on human-produced data. The result is a tight coupling between large-scale user numbers, large-scale data, and large-scale models which is understood to produce runaway feedback loops for a tech company and their proprietary algorithms \cite{wong2023}. This triple coupling is highly unusual in, for example, meteorology's big data system -- more users will not usually improve a weather app's predictions -- and its achievement in AI/ML has required the use of specific data types (internet/platform data) and specific monetization practices (information technology) that distinguishes the AI/ML regime of scale in particular ways compared to other data organizing strategies.

We center our findings on frictions between state and entrepreneurial regimes of scale at three different points in the data pipeline: observations, data, and models. First, regimes of scale operate at the level of \textit{observations}, the ways that observations are embedded into large-scale networks for downstream use. Under state scale, observations that are further from the ground (e.g. balloons collecting pressure) work better for meteorology's global-scale foundation models, while ground observations work better for finetuning models to specific weather applications. This hierarchy of observations breaks with AI/ML observational scale, which relies on growing and monitoring dense networks of human users and their generated data. Second, regimes of scale operate at the level of \textit{data}, the global-scale datasets that are produced from observations and which serve as the initial conditions for weather/climate models. Initial AI/ML weather models utilized these datasets but were hindered by the physical laws governing their composition. This is set to change as AI/ML practitioners bypass these datasets and the centralized state agencies that produce them, embracing a new entrepreneurial network of remote sensing data infrastructures and retailers. Finally, meteorological \textit{models} are able to make state-scale claims that are transdisciplinary and transcendent because their forecasts are used across domains like water management, forest management, insurance, and finance. AI/ML weather models have faced challenges in moving across disciplines in this way because they break fundamental physical laws.

Through mapping out these frictions across two data pipelines that are increasingly intertwined with one another, this paper contributes to HCI in two ways:
\begin{itemize}
    \item First, we demonstrate that the framing of AI/ML as domain agnostic is insufficient when it comes to studying AI/ML applications. Our empirical study of AI/ML's integration into meteorology, itself a Big Data system, shows that many AI/ML frictions come from conflicting ways of organizing Big Data rather than "just" a conflict between transcendental AI/ML techniques and situated domain knowledge. We demonstrate that AI/ML and physics-based meteorology have distinct ways of scaling observations, data, and models, and that these conflicting ways of scaling produce many of the frictions we see in AI/ML in practice. 
    \item Second, we develop a framework, \textit{regimes of scale}, to track fundamentally different ways of understanding how observations/data/models could or should be scaled up. We compare the entrepreneurial scale of AI/ML, which relies on a close relationship between user numbers, derived data, and model power, with the state scale of meteorology, which relies on precision-nesting local or national observations into global-scale datasets. Rather than there being a single path to scale, we show that features which make AI/ML data, observations, and models scalable can be actively detrimental to scale as it is organized in meteorology.
    
\end{itemize}

The rest of this paper is structured as follows: First, we provide a definition of our regimes of scale framework [2.1], a comprehensive background of HCI research on AI applications [2.2], and a review of both regimes of scale (meteorology [2.3] and AI/ML [2.4]). Next, we detail the methods of our interview study [3]. We then map frictions between the AI/ML and meteorological regimes of scale at three points in the meteorological data pipeline: observations [4.1], data [4.2], and models [4.3]. Finally, in our discussion and conclusion, we discuss two strengths of the "regimes of scale" framework: in understanding AI/ML applications relationally with respect to a domain's existing Big Data pipeline [5.1], and in understanding AI/ML applications with respect to a target domain's existing interdisciplinary network of users [5.2]. Fundamentally, this paper aims to intervene in HCI by foregrounding how AI/ML methods are reckoning with well-established and fundamentally different ways of organizing large-scale data. As AI-first initiatives attempt to entirely restructure domain institutions, the individual or group "domain expert" may be less crucial to consider than the well-rooted data infrastructures that AI/ML methods are now attempting to displace. 

\section{Background}

\subsection{Regimes of Scale: Definition}

Our paper's primary contribution is what we call "regimes of scale," by which we mean a given technoscience's strategy of building and maintaining data infrastructures at scale. This framework builds on Thomas Kuhn's notion of scientific paradigms \cite{kuhn1970} and Michel Foucault's notion of "regimes of truth" \cite{foucault1976}. Namely, we draw from Kuhn's notion of scientific paradigms, which centers theory and the professional development of individual scientists: as Kuhn argues, a paradigm determines the pattern of "normal science" in a given era before facing enough internal contradictions to force a paradigm shift to an entirely new way of creating scientific truth \cite{kuhn1970}. Foucault expanded this analysis with "regimes of truth", which considers institutions and the institutional disciplining of truth; rather than Kuhn's question of theoretical coherence, Foucault's regimes of truth hones in on how truth is governed and how facts are institutionally verified in a technoscientific system \cite{foucault1976}. For Foucault this was primarily an institutional/discursive question, "a question of what governs statements, the way they govern each other," and a question of who is permitted to define and enforce truth. These institutional questions are a major part of this paper's analysis. US meteorology is tightly defined by the institutional practices, strategies, and norms of NOAA (the US state meteorological agency), and NOAA plays a large part in deciding what makes good or bad meteorological data in and outside the US. More importantly, though, we show in our paper how "regimes of scale" comprise not just the institutional disciplining of meteorological truth but also the organization, maintenance, and stabilization of data, observations, and models within AI/ML and meteorological data infrastructures.

\textbf{If regimes of truth determine what is true and how things become true, regimes of scale determine how things become scalable, how scale can be achieved (at what cost), and what scale itself means for a given data system.} Our findings demonstrate that the entrepreneurial scale of AI/ML and the state scale of meteorological data operations differ not only at the theoretical level but also in the institutional and infrastructural relationships that allow data, observations, and models to scale in each field. This appears in the institutional networks and expertise required to produce relevant data, observations, and models in each field; it appears in the degree to which public and private entities are licensed to scale these data, observations, and models; and it appears in the ways in which data, observations, and models are judged to be good or bad as they attempt to cross boundaries within a regime of scale's network of users. By shifting the focus of our analysis to regimes of scale, we can productively understand how meteorology and AI/ML scale differently and how this can produce fundamental frictions between both fields beyond just AI/ML's mismatch of/with domain knowledge.


\subsection{HCI Literature Review: Application Domain, Knowledge Domain, and Infrastructure for Domains}

This work contributes to the HCI literature on AI/ML and data-scientific interactions with domains, which can be understood in three different subsections: domains as application points, as sources of expert knowledge, and as end users for AI/ML methods. The most straightforward use of domain in HCI is as an "application domain," which typically ties a wide range of different domains into an overarching study of AI/ML's integration into existing data pipelines \cite{gurita2025,yildirim2023,kim2024,shin2025,yang2020}. For example, Lee et al. test the applicability of card-based AI/ML model explanations across the application domains of healthcare, finance, and management \cite{lee2024}. Likewise, Sambasivan et al.'s study of high-stakes domains discusses healthcare, food/agriculture, environment/climate, and more to study data cascades via interviews of AI practitioners \cite{sambasivan2021}. Even within specific domains like healthcare, application domain can be used to work across different subdomains: for example, Thakkar et. al use "application domain" to split public health into project domains including maternal and sexual health \cite{thakkar2022}. A consequence of terms such as application domain is that very specific groups are regarded as being domain-agnostic: data scientists and data managers in Thakkar et al.'s study see themselves as being able to work across many domains or subdomains in their work, and this centers data scientists as having the exclusive ability to traverse data pipelines. Per Jung et. al, this framing of data experts as "domain agnostic" can yield research that unwittingly frames data scientists as having the ability to develop statistical models or algorithms, "with domain experts limited to the role of informers" \cite{jung2024}. 

Efforts to address this instrumentalization of domain experts have leaned on another use of domain: domain as "domain knowledge" \cite{calota2025}. Domain knowledge is fundamentally tied to the figure of the domain expert who holds it, and the term has been used in HCI to advocate for further participation of domain experts in data science work \cite{alvaradogarcia2025}. The precise benefit of domain expert participation varies. For Jung et. al it improves the actionability of data science systems \cite{jung2024}, for Bhattacharya et. al it can help decrease representation bias \cite{bhattacharya2025}, and for Sambasivan et. al acknowledging domain expertise is simultaneously an ethical question (given the deskilling of domain expertise in low-resource areas) and a practice which improves AI engineering fundamentals like getting good quality data and getting timely feedback on deployment \cite{sambasivan2022}. Finally, Freeman et. al study domain experts' answering strategies in order to use domain expertise as a resource for making LLMs useful even for users unskilled at a target domain.

Other HCI works use the term "domain expert" to study interactions between data scientists and domain experts in organizational settings, or to study interactions between domain experts and data-scientific tools as users \cite{burton2012,lee2024,zhong2025}. Mao et. al study teams of bio-medical scientists collaborating with data scientists, and frame their work in terms of interdisciplinary collaborations; bio-medical and data scientists had to find common ground to effectively transform domain-specific biomedical research questions into computable data science questions \cite{mao2019}. Zhang et. al more broadly consider collaborations between different types of data science workers (engineer, manager, researcher, domain expert, communicator), arguing that domain experts are active at every stage, but take on particularly prominent roles when evaluating and communicating results \cite{zhang2022}. A separate branch of HCI studies domain experts as \textit{users} of data-scientific tools: Ziegler and Chasins study users of geospatial data \cite{ziegler2023}, Jung et. al study craft brewers using brew sheet data \cite{jung2022}, and Lakier et. al study marine scientists who use data science tools but do not generally consider themselves data scientists \cite{lakier2025}. Finally, certain works understand data and data-scientific tools as infrastructure for domains. This was most commonly seen in the cyberinfrastructure studies of the 2000s within CSCW and HCI scholarship \cite{ribes2009a,baker2005}, which often aimed to create or study domain infrastructures at scale \cite{ribes2014,ribes2014a}, and it continues today: Neang et. al study oceanographic infrastructure \cite{neang2021,neang2023}, and Steindhart and Jackson likewise study collaborative ocean science \cite{steinhardt2014a}. 

This paper studies interactions between AI/ML and meteorology via an interview study of 18 forecasters/meteorologists, 8 of whom are also AI/ML practitioners. Meteorology acts as a domain for AI/ML application, but it also operates as a Big Data regime with application domains of its own: meteorological predictions are used in disaster management, energy, insurance, and finance, for example. Additionally, just like bio-medical scientists \cite{mao2019} and marine scientists \cite{lakier2025}, meteorologists use data-scientific tools and methods to manage their own data and to spread it across disciplines. This means that the relationship between AI/ML and meteorology cannot be readily captured by the three types of relationships -- extractive, collaborative, and invisible -- that domain science has with AI/ML. Understanding meteorology as a distinct technoscience with its own particular ways of scaling data infrastructures, we intervene in studies of "domain applications" by advancing a framework, regimes of scale, which can trace interactions between different interdisciplinary Big Data infrastructures.

\subsection{Meteorological Big Data and the US Weather Enterprise}

Our object of study is the meteorological Big Data regime, specifically the US Weather, Water, and Climate Enterprise, a public-academic-private meteorological assemblage formed via a series of institutional agreements over the 1990s. Since meteorology's rapid professionalization in the 1960s and 70s concomitant with the rise of general circulation models (GCMs), the field has operated as a kind of national-global science -- each country has their own general circulation model, and the national weather agencies (e.g. NOAA in the US, the IMD in India, the CMA in China, the Met Office in Britain) serve central roles in the management of observation networks and the provision of downstream weather services for each country. We call this regime "state scale" because it relies on a national and international process of nesting meteorological networks and their accompanying territorial claims into a world model.

Meteorology since the 1970s has rested on GCMs, massive physics-based models which simulate the circulation of air on a global scale \cite{edwards2013}. Models use data assimilation to pull from a wide range of meteorological observations, from ground stations to radiosonde (balloon) observations to satellite soundings to marine buoy networks \cite{edwards2013}; these observations are then denoised, translated, and combined through the use of other observations or laws under a common assumption of physical consistency, simulating evenly spaced and global reanalysis data such as the ERA-5 dataset \cite{pu2018}. GCMs predict future weather and climate from these reanalysis datasets, and running them at a high enough resolution to effectively model weather phenomena is extremely computationally expensive; the emergence of GCMs in the 1960s hinged on the availability of supercomputing power \cite{dalmedico2001} and the global vision of satellites \cite{ramage1971}. Physics-based weather modeling therefore emerged in the context of newly massive data (satellite) and compute infrastructures, and model accuracy has improved incrementally over the last fifty years via a "quiet revolution" of Moore's-law computing advances and systematic efforts to mine sources of predictability \cite{dalmedico2001}. State-of-the-art models take hours to run even on the weather agencies' massive supercomputing infrastructures; this high CPU requirement has traditionally made it difficult to impossible for private companies to run GCMs for their own use. All of this yields a kind of infrastructural globalism, per Paul Edwards \cite{edwards2006}, by which world organizations like the WMO mediate data while national meteorological agencies (NOAA, IMD, CMA) organize their own observation, modeling, and forecasting networks. 

\subsection{AI/ML Big Data and Foundation Models}

Unlike meteorological Big Data, AI/ML data/model infrastructures cannot be cleanly located in national and international state institutions; instead, they are located in multinational tech corporations who control AI/ML software \cite{widder2023}, research and development \cite{ahmed2023}, infrastructural GPU compute \cite{schmid2024}, and cloud computing services \cite{narayan2022}. This control of AI/ML technology in a few corporations, combined with what Devika Narayan calls the "new logics of scaling" enabled by cloud computing infrastructures \cite{narayan2022}, yields a regime of scale we call \textit{entrepreneurial scale} due to the close intertwinement of large-scale user numbers, large-scale data, and large-scale models in tech companies. Entrepreneurial scale allows these companies to anticipate what Jamie Wong calls "runaway feedback loops" as increasing user counts increase the amount of proprietary data available to a company \cite{seaver2021}, compounding on itself towards an explosive and transformative change in a company and its proprietary models \cite{wong2023}.

This political-economic relationship is embedded in the theory and practice of modern AI/ML methods. While AI as a term traces back to the 1960s \cite{minsky1961}, its contemporary methods were developed in the 1980s and 1990s before emerging spectacularly in the late 2000s. This emergence of what we call modern AI/ML is usually traced to two data challenges: the 2009 Netflix Challenge, which tasked competitors with effectively recommending movies to users \cite{zhang2024}, and the 2012 Imagenet image identification challenge \cite{whittaker2021}. In each case, AI/ML methods outperformed conventional statistics due to their ability to exploit modern data/compute infrastructures, in particular the surge of data/compute from platform tech companies, the Internet, and the compute scaling of Moore's law \cite{zhang2024}. Modern AI/ML theories such as emergence \cite{wei2022} and model architectures such as the Transformer \cite{vaswani2023} prioritize exploiting massive human-produced data (with an implicit "scale" measured in users) in anticipation of fundamental paradigm shifts in a model. This is a highly specific way to scale, one currently unavailable to meteorological Big Data that is dependent on intentional observations and a slow progression of weather modeling \cite{bauer2015}. As our Findings show, to many meteorologists, AI/ML methods themselves are closely tied to the potential ability to scale like a tech company.

\section{Methods}

We have established this paper's two regimes of scale: the state scale of national meteorology and the entrepreneurial scale of modern AI/ML methods. The first significant contact between these regimes of scale occurred in 2022, when a wave of AI weather models were created by skilled AI/ML research groups in large tech companies: FourCastNet from NVIDIA \cite{pathak2022}, Pangu-Weather from Huawei Cloud \cite{bi2023}, and GraphCast from Google DeepMind \cite{lam2023}. Many of these AI/ML teams did not have a meteorologist in the room; weather was essentially a time series data problem. Due to this institutional gap, meteorological interest in AI/ML modeling progressed slowly over 2023, with a turning point per A1's fieldwork being the American Meteorological Society conference in January 2024. This paper's interviews, conducted in summer 2025 and 2026, took place during an ongoing push to create integrated meteorologist-data-scientist teams, and, simultaneously, a push to create direct-from-observations AI/ML weather models which fully bypassed the meteorological data pipeline. 

This paper draws from 18 30-60 minute interviews conducted by A1 with public, private, and academic meteorologists in the United States. Over 2025 and 2026, A1 built a list of 42 relevant actors through their attendance at two community meetings attended by forecasters, public/private meteorologists, and downstream users of weather data. These actors were purposefully selected for their exposure to AI/ML methods as part of their work. Of the 18 final participants in the study, 15 had direct exposure to AI/ML methods in their daily work, either as AI/ML practitioners themselves or as corporate or state users of AI/ML model outputs. The remaining three participants helped provide perspective on aspects of the meteorological data pipeline that had been less directly engaged by AI/ML, such as climate science and county/state-level forecasting. The participants' general institutions and roles are as follows: 

\begin{table}[]
    \centering
    \begin{tabular}{c|c|c|c|c}
         - & General institution & Role & Degree & Years of experience \\
         P1 & Tech company & Head meteorologist & Ph.D & 30-40\\
         P2 & US agency & Head scientist & Ph.D [geology] & 30-40\\
         P3 & University & Research meteorologist & Ph.D & 20-30\\
         P4 & Private weather company & CEO & Ph.D & 10-20\\
         P5 & Private weather company & Project manager & M.S. & 10-20\\
         P6 & AI weather startup & CEO & B.A. [philosophy] & 10-20\\
         P7 & NOAA/NWS & Meteorologist In Charge & B.S. & 20-30\\
         P8 & NOAA/NWS & Research meteorologist (AI/ML) & Ph.D & 10-20\\
         P9 & Private weather company & Head meteorologist & Ph.D & 30-40\\
         P10 & Private weather company & Head meteorologist & Ph.D & 40-50\\
         P11 & NOAA/NWS & Research meteorologist (AI/ML) & Ph.D & 10-20\\
         P12 & AI weather startup & Head meteorologist & Ph.D & 10-20\\
         P13 & Private weather company & Head meteorologist & Ph.D & 30-40\\
         P14 & AI weather startup & Head meteorologist & Ph.D & 20-30\\
         P15 & University & Research meteorologist & Ph.D & 10-20\\
         P16 & Weather policy company & Policy expert & M.P. [Policy] & 20-30 \\
         P17 & Private weather company & Head meteorologist & Ph.D & 10-20\\
         P18 & NOAA/NWS & Project manager [former] & M.S. [Civil eng.] & 50-60\\
    \end{tabular}
    \caption{Participants and their general institutions and roles. All degrees are in meteorology unless otherwise stated.}
    \label{tab:placeholder}
\end{table}


The interviews were systematically coded over 2025 and 2026 in two main stages. First, A1 coded the interviews according to the type and subtype of the meteorological objects being discussed (observations, data, models), any application domains being discussed (insurance, finance, disaster support), and the ownership of the meteorological objects being discussed (national, private, state). Simultaneously, A1 coded for participants' discussion of AI/ML methods, coding around the affordances of AI/ML to interviewees, the concerns interviewees had with AI/ML methods, and the demands AI/ML methods made of meteorological infrastructures and organizations in order to provide value. In the second stage, and after weekly discussion between A1 and A2, we created subcodes in each of the initial categories (such as Observations/"are the king" and Private weather/"differentiation") and also added a main code, "qualities in translation," intended to track terms that were used differently across AI/ML and meteorology. Our developed paper is organized around \textbf{scale}, a term which serves as a marker of success for both AI/ML and meteorological Big Data systems, but in significantly different ways

\subsection{Positionality Statement}

Our team consists of A1, a graduate student, and A2, a faculty member at the same R1 academic institution in the United States. The first author is a US social scientist trained in AI/ML statistical theory who has interfaced with meteorologists in and outside of the Global North since 2023; her US citizen status played a significant role in studying and interviewing the unusually privatized (for meteorology) weather-water-climate assemblage in the United States. The second author is a US-based ethnographer who has spent more than a decade studying environmental data infrastructures and governance in Southeast Asia and the U.S., allowing her to engage with environmental and earth scientists, including meteorologists and remote sensing scientists, who have attempted to make their datasets more AI- and ML-ready. At the same time, she acknowledges that her longstanding ethnographic experience with such experts stems largely from a particular moment in time (2015 - 2022) where Generative AI and fears of displacement were less prevalent. As such, together with A1, we aimed to hold both ethnographic and fieldwork periods side by side, examining the tensions, alignments, and contradictions that arose from analyzing the interviews together.

\section{Findings}

\begin{table}[]
    \centering
    \begin{tabular}{c|p{35mm}|p{40mm}|p{35mm}}
         - & Meteorology (State scale) & AI/ML (Entrepreneurial scale) & AI/ML Affordance for Meteorology \\\hline
         Observations & Nesting observations into global scale & Recursive scaling from user obs & Obs as "gold" (inherently valuable)\\\hline
         Data & Generated at scale using physics models & Preprocessed at scale via labeling / other models& Use of indirect (satellite/radar) data\\\hline
         Models & Claim scale via representation of large-scale patterns & Claim scale by improving with massive compute/data & Faster models\\
    \end{tabular}
    \caption{How observations, data, and models claim or fit into scale in the meteorological and AI/ML regimes.}
    \label{tab:placeholder}
\end{table}


AI/ML methods are not new to meteorology. Because of global climate's reputation as a highly chaotic system, researchers had tested neural networks for weather/climate prediction as early as the 1990s \cite{elsner1992}, and the 2000s and 2010s witnessed various applications of modern AI/ML methods as powerful semi-statistical tools suitable for the nonlinear patterns of weather/climate \cite{kumar2012,singh2013,dibike2006,wang2018}. Because of the historical ties between meteorology and AI/ML methods, and because traditional meteorology is already a field dealing with large and complex datasets (what we call Big Data), many participants framed the question of trusting/using this new wave of AI/ML models in terms of prior technologies/datasets that have been integrated into the field of meteorology. P9, the head meteorologist at a private weather observation company, compared AI evaluation to the evaluation of novel satellite sensors:

\begin{quote}
    We need an underwriter's lab for AI evaluation, but we need the same thing with C9 [a satellite observation company], who has these satellite based microwave sounders... [P9]
\end{quote}


\begin{figure}
    \centering
    \includegraphics[width=1\linewidth]{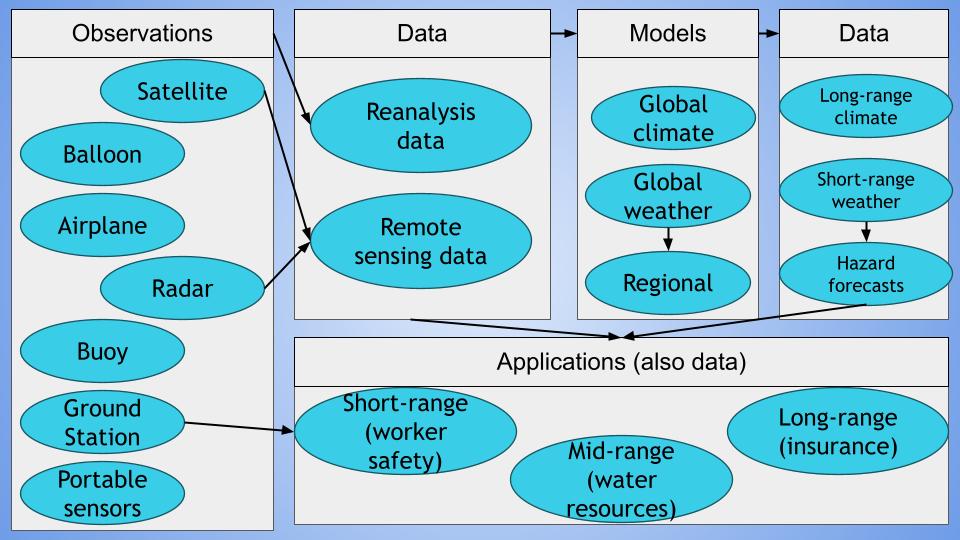}
    \caption{The meteorological "data pipeline," loosely split into observations, data, and models.}
    \Description[The meteorological "data pipeline," loosely split into observations, data, and models.]{Four boxes titled Observations, Data, Models, and Data. The Observations box contains five seven bubbles: Satellite, Balloon, Airplane, Radar, Buoy, Ground Station, and Portable Sensors. The Data box contains Reanalysis data and Remote sensing data. The Models box contains Global climate, Global weather, and Regional bubbles, stacked vertically, with each bubble pointing to the one below it. The second Data box contains Long-range climate, Short-range weather, and Hazard bubbles, representing different kinds of data products produced by models. Finally, there is an "Applications (also data)" box spanning the bottom of the figure, containing Short-range (worker safety), Mid-range (water resources), and Long-range (insurance) applications. Some observations are connected directly to the "short-range" bubble, and both "data" boxes are connected to the "Applications" box.}
    \label{fig:pipeline}
\end{figure}


However, meteorologist interviewees also noticed a severe professional divide from AI/ML methods compared to other then-novel technologies, particularly due to the private research lab structure of AI/ML research [P17], which (per A1's fieldwork at meteorological conferences) made correspondence difficult during 2022 and 2023. As this professional divide has been bridged over 2024 and 2025, particularly with the rise of mixed teams of meteorologists and AI/ML researchers in Nvidia and the ECMWF, meteorologists have been increasingly confronted by fundamental differences in how AI/ML observations, data, and models functioned, which appeared both as frictions and as potential affordances of AI weather models. 

We organize our findings around these affordances and frictions at three sites in the meteorological data pipeline: observations, data, and models. Meteorological \textbf{data} refers to the inferred parameters of satellite observation systems and other remote sensing tools, the reanalysis data produced by assimilating observations via physics-based models, and the long-term climatology data produced by massive models \cite{emanuel2020}. \textbf{Observations} are less ambiguous -- they usually correspond to direct records of meteorological metrics like precipitation, temperature, and humidity, although they have been lent to less direct measurements like satellite and radar methods. Even so, several interviewees noted that more indirect observations (like satellites and radar) were "basically models" [P4,P8,P12,P14], not exactly observations in the same way as direct measurements. This makes observations operate similarly to ground truth in AI/ML: the term has a strong association with physical realness. Finally, \textbf{models} in the US meteorological context almost exclusively refer to General Circulation Models (GCMs), massive physics-based models of the global atmosphere that take million-dollar supercomputers to run. These are vaguely analogous to the foundation models of AI/ML, large and expensive blocks of compute and data that form a foundation for downstream research. We split our findings across data, observations, and models to reveal fundamental gaps between the state scale of meteorology and the entrepreneurial scale of AI/ML.
\subsection{Scaling Observations: Predictability in Air/Ground Observations}

\begin{figure}
    \centering
    \includegraphics[width=1\linewidth]{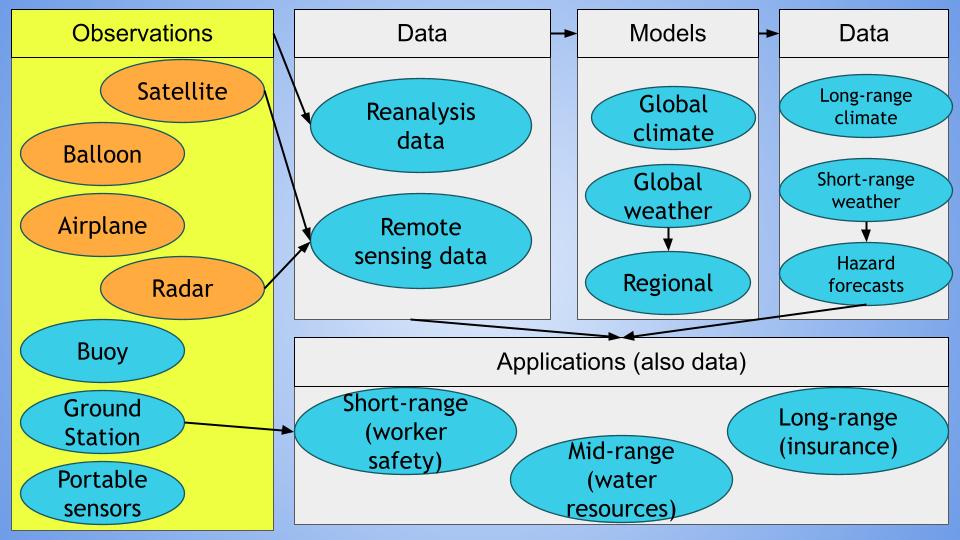}
    \caption{The current section discusses \textbf{observations} (highlighted). The observations of the free atmosphere are marked in orange.}
    \Description[The meteorological "data pipeline," loosely split into observations, data, and models. The Observations box is highlighted.]{Four boxes titled Observations, Data, Models, and Data. The Observations box contains five seven bubbles: Satellite, Balloon, Airplane, Radar, Buoy, Ground Station, and Portable Sensors. The Data box contains Reanalysis data and Remote sensing data. The Models box contains Global climate, Global weather, and Regional bubbles, stacked vertically, with each bubble pointing to the one below it. The second Data box contains Long-range climate, Short-range weather, and Hazard bubbles, representing different kinds of data products produced by models. Finally, there is an "Applications (also data)" box spanning the bottom of the figure, containing Short-range (worker safety), Mid-range (water resources), and Long-range (insurance) applications. Some observations are connected directly to the "short-range" bubble, and both "data" boxes are connected to the "Applications" box. The Observations box is highlighted.}
    \label{fig:pipeline_obs}
\end{figure}


\begin{figure}
    \centering
    \includegraphics[width=1\linewidth]{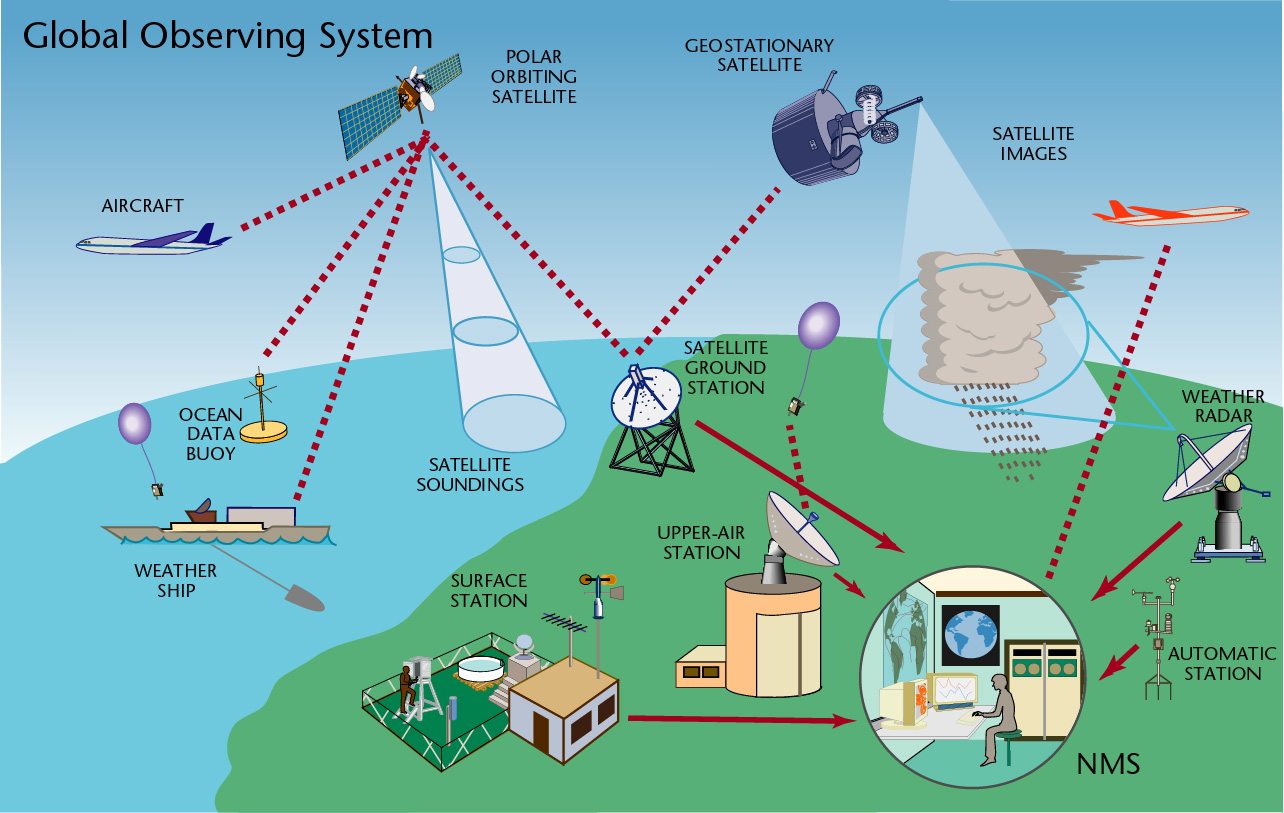}
    \caption{Some observations measure the upper air, while others measure the surface. From WMO, "Observation components of the Global Observing System"}
    \Description{An educational figure of various types of meteorological observations. Some observations (satellite, aircraft, radiosonde) sit in the upper air, while others (buoy, ground, weather ship) sit on the surface.}
    \label{fig:WMO_obs}
\end{figure}

\subsubsection{Observation Types in the Meteorological Data Pipeline}

Modern atmospheric science categorizes observations into two types of observations: ground observations that are associated with verification and small-scale model finetuning, and upper air observations that are associated with large-scale models. This is because air near the ground is dominated by friction and local topography, while air further above the ground, the "free atmosphere," is dominated by pressure winds and general circulation \cite{stull2012}. As such, observation types in meteorology roughly correspond to a smaller/larger spatial scale: ground observations are appropriate for small-scale work and model finetuning, while upper-air observations (aircraft, balloon, satellite) provide the information necessary to feed the field's massive general circulation models (GCMs). P12, a machine learning researcher and meteorologist, described this difference between ground and upper-air observations as a difference in information quantity: with observations near the ground "it seems like it's a lot of information because you can have very dense networks, very high frequency and time," but counterintuitively, "you're just generally not measuring information that's helpful for the [global] models" [P12]. The basic assumptions of atmospheric physics mean that ground observations, even a massive number of ground observations, cannot effectively define the atmosphere on a global scale, marking these ground observations as intrinsically smaller-scale. But ground observations are still critical for verifying forecasts [P1] and fine-tuning them to local or domain-specific contexts. Per P4, a head meteorologist at a weather services company, "there's no better hyper local [prediction/forecast] without a sensor on site!" As such, the distinction between ground and upper-air observations marks them as smaller or larger scale respectively, but also determines where these observation types are best used in the meteorological data pipeline.

Just as crucially, scale also determines how observations are distributed across public and private entities. As P4, a CEO of a private weather company, described:

\begin{quote}
    The foundation of weather modeling is all the observational infrastructure, then it's the computational power. Then it's data analytics, and then data dissemination. And historically, I used to plot all this out when I'm talking to investors, historically the private sector would would live on the top of that, basically doing the dissemination, and doing more than just disseminating data, but like doing further analytics, further down, scaling, further optimization and derivations for customers. And that's where the private sector has usually lived. [P4]
\end{quote}

The above quote shows how the private sector is primarily responsible for deriving new value from ground observations through further analysis, customization, and distribution to clients. Foundational upper-air observations, in contrast, are overwhelmingly produced by state-affiliated or public institutions: per P6, "the upper air stations, the balloon stations, those are almost entirely government owned and operated. You do have universities [who] will do research flights, or some companies will do research flights, but that usually doesn't get assimilated or even archived anywhere, really." As of the time of writing, four major weather/climate companies in the US produce upper-air and global-scale observations for NOAA -- Climavision (radar observations), Windborne (balloon observations), Tomorrow.io (satellite observations), and Spire (radio occultations). For the rest of the private weather companies in the US, research participants noted that they typically utilized more local forms of proprietary data -- the company's own ocean or ground sensors, or the local sensor networks of client institutions. These companies' "value add" was through local data or through domain-specific adaptations of NOAA's public offerings.

These ground observations, largely produced by the private sector, faced additional barriers to scaling up due to their entanglement with the local needs of client institutions. P1, the head meteorologist at a tech company, told us that client institutions often placed sensors in ways that were inherently difficult to scale -- "for example, in road weather, we sold a lot of stations to road weather companies and agencies, and they only put the weather stations where the problem places were on the road, exactly... if they installed them in a grid pattern, like one every couple miles, something like that, you could ingest that into a model, and that helps your forecast model." Because of the growing demand for adapting observation data for different functionalities such as road/forest weather prediction, ground observations were in a sense made less scalable than upper-air observations, both because of physical law and because those observation networks are entangled with the smaller-scale demands of local or domain-specific institutions. 

Still, the problem of local ground measurements could be bypassed by the few large observation companies that maintain global upper-air observation networks, but this revealed another major roadblock to private global-scale predictions -- they struggled to differentiate themselves from publicly-available forecasts. Over the last fifty years, meteorological models have improved over a slow "quiet revolution" as observation and model infrastructures have grown incrementally more robust \cite{bauer2015}. The current weather/climate observation infrastructure, formed through the state scale of meteorology, is the product of hundreds of institutional actors combining data into a few nationally-produced global models that are relatively cheap to access. This has made it extremely difficult for private companies to deliver global-scale weather/climate predictions that were meaningfully better than readily accessible, publicly-produced baselines like the Global Forecast System [GFS] hosted by the US National Weather Service (NWS). 

As compute became cheaper over the mid to late 2010s, all four of the global met observation companies attempted to create their own proprietary weather/climate models; but as several interviewees noted [P4,P9,P10,P12] these models faced extreme difficulty significantly improving on or even matching the massive state models developed by NOAA and delivered by the NWS. Per P9, "It turns out that that not everyone was going to flock to them to buy forecasts. You know, especially in the United States, NOAA puts out a pretty damn good product for free." Because the data, expertise, and compute necessary to build foundational GCMs are all so heavily centralized in the national weather agencies under what we call state scale, even robust global-scale observations developed by the private industry could not yield global-scale forecasts that could compete with the free or subsidized products of government agencies. Instead, private observation companies were forced to sell their observations to NOAA and other national agencies [P10], inherently capping the fungibility or value of large-scale weather observations. In P4's words, "that's the problem with these data purchases -- you basically have one customer per country." While this is certainly a problem for private observation companies, this nonfungibility of observations is arguably what makes state scale function, as it allows national agencies to efficiently centralize observations and scale them into a global observation infrastructure.

\subsubsection{AI/ML's transformation of observations' value}


The section above demonstrates how the scale or scalability of observations maps onto their type (upper-air vs ground), their perceived role (foundational vs last-mile), and their institutional home (public vs private). In this context, private weather companies face two major obstacles to achieving valuable prediction at a large scale: private observations largely consist of local-scale ground observations (where the people are), and even robust global observation networks cannot easily be translated into profit without the foundational expertise of the national agencies. Against this backdrop, AI/ML methods have promised to make ground observations more valuable and to make observations more easily translate into meaningful forecasts. The rest of this section lays out these shifts and demonstrates that they correspond to AI/ML's entrepreneurial regime of scale.


According to some research participants, one promise of the new AI/ML weather/climate models is their potential to make ground observations work for global-scale models. Traditionally, observations in meteorology are incorporated into models via data assimilation, which interprets them through the lens of a physics-based model. Because these models operate under boundary layer assumptions, ground observations naturally travel less than upper-air observations, and so they are inherently smaller-scale. But with an AI/ML model, because "we don't define the atmosphere in terms of physical components like we would with Navier-Stokes" [P12], it may be possible to make ground observations meaningful at a global scale. P12 argued that because AI/ML models are detached from traditional physics-based laws of what data can carry information at global scale, they may be able to make use of massive (but currently fundamentally local-scale) ground data sets. 

This potential application of ground observations at a global scale is particularly appealing for private weather companies with institutional clients. These clients are typically interested in weather as a relevant factor in other domains like construction or financial modeling or water resource management, and so ground observations more easily ride on their existing corporate infrastructure. As noted by \cite{zhang2024}, AI/ML methods became viable due to the Internet and the proliferation of platform data. This data is free in the sense that it extracts large datasets from the everyday lives and behaviors of users \textit{without their active consent}. Global scale meteorological observations, and even traditional ground observations, lack this sort of free generation: they cannot easily appropriate the work of users, and instead, per P1, "you have to say, hey, can I put a 60 foot concrete pole in your front yard?" [P1]. In this context, the ability to make use of existing ground datasets seemed quite valuable. For example (per P12), "If you have wind measurements from the cells of turbines, why not feed that into a sufficiently high resolution model to provide more context?" As P12 noted, this also stood to change \textit{how observations appear with respect to data/models} -- "in traditional modeling, we treat the observation system as static," but incorporating new forms of data such as platform-generated data could create what P12 notes as a more mobile and flexible observation infrastructure, directly linking "the observation system that we deploy in the real world and the forecast that comes out of it." The promises of AI/ML lies in its perceived ability to transform what constitutes scalable observation data in the first place. Instead of relying solely on physical observation stations on the ground, user-generated data from sensors located in other infrastructures are sourced as a kind of observation data that could be used for forecasting and serving institutional and private clients. AI/ML's entrepreneurial scale, instantiated through the triple relationship between user numbers, data, and models, is introduced in meteorology by changing how observations scale and what they can do. This transformation is deeply contingent on the private industry's longstanding role as a provider of ground observations.



Simultaneously, AI/ML methods are often accompanied by the rhetoric that more or better observations will translate directly into profitable forecasting improvements. In contrast to meteorology's state scaling via an incremental quiet revolution (see Section 4.1.1), the entrepreneurial scale of AI/ML (see Section 2.4) promises recursive and transformative improvements to models and their host companies with improved observation quantity/quality, drawing a much more direct link between better and more observations and competitive advantage. P6 put this most explicitly: "The big thing that seemed clear to us, based on how other AI tech has progressed, was that data will quickly become more important than the algorithmic improvements. Whether that's more data, or better data, or better labeled data." Similarly, per P4, "there are some private companies now that are kind of picking up on that and saying, well, I have all the satellite data, so now I have the currency of machine learning." Under the marketing rhetoric of AI/ML models exhibiting emergent abilities and undergoing transformative change through increasing access to data and compute, observations suddenly seemed to be agnostically useful and highly fungible. This entrepreneurial regime of scale is envisioned by interviewees as promising to yield more and better observations for global-scale predictions within a single vertically integrated company, making both ground and upper-air observations increasingly valuable outside of NOAA's centralized data infrastructures.

The current attempt to inject raw observations with explosive power has some precedence in meteorology: several interviewees [P1,P8,P9] noted that this script of data-as-gold had been employed by various private companies with different technologies (airplane observations, radio occulatations) over the last three decades, ever since the partial liberalization of US meteorology in the 1990s. But the AI/ML regime of scale now appears as a new and undeniably successful way to make use of observations in the fungible and explosive ways observed in tech companies. Notably, AI/ML methods have also caused meteorological data to be more fungible because they are increasingly valuable to non-meteorological actors. P5, who worked in a data aggregation company, noted an uptick in "more AI, machine learning kind of people" as customers. Ultimately, AI/ML ways of scaling promised to make observations fungible, to enable private infrastructures of global observation networks, and to dissolve traditional meteorological boundaries between ground and upper-air observations. In both technical and non-technical contexts, we see that the affordances of AI/ML methods for meteorology are understood not just in terms of raw theory/method, but in terms of strategies of organizing data infrastructures that are conventionally associated with tech companies.

\subsection{Scaling Data: Satellite Data and Reanalysis Data}

\begin{figure}
    \centering
    \includegraphics[width=1\linewidth]{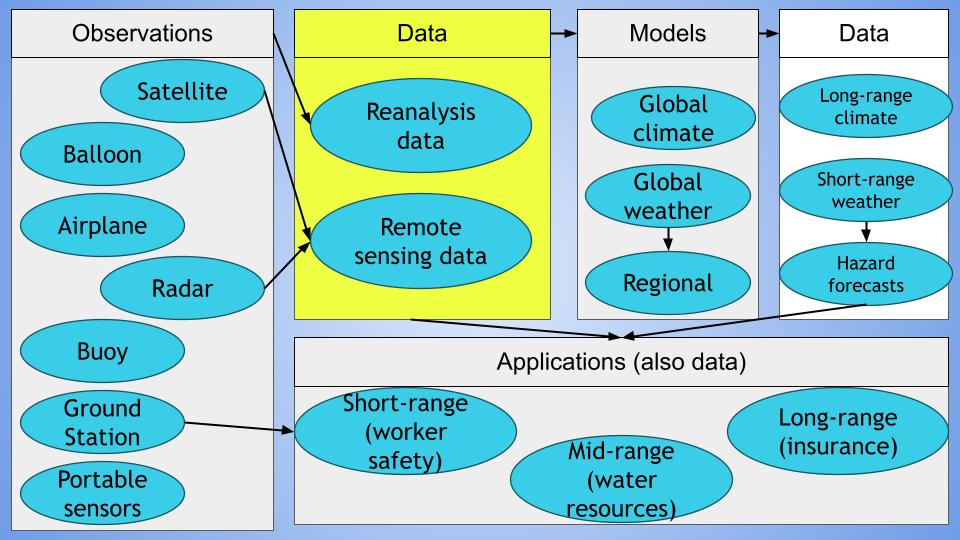}
    \caption{The current section discusses \textbf{reanalysis and remote sensing data}}
    \Description[The meteorological "data pipeline," loosely split into observations, data, and models. The Data box is highlighted.]{Four boxes titled Observations, Data, Models, and Data. The Observations box contains five seven bubbles: Satellite, Balloon, Airplane, Radar, Buoy, Ground Station, and Portable Sensors. The Data box contains Reanalysis data and Remote sensing data. The Models box contains Global climate, Global weather, and Regional bubbles, stacked vertically, with each bubble pointing to the one below it. The second Data box contains Long-range climate, Short-range weather, and Hazard bubbles, representing different kinds of data products produced by models. Finally, there is an "Applications (also data)" box spanning the bottom of the figure, containing Short-range (worker safety), Mid-range (water resources), and Long-range (insurance) applications. Some observations are connected directly to the "short-range" bubble, and both "data" boxes are connected to the "Applications" box. The Data box is highlighted.}
    \label{fig:pipeline_reanalysis}
\end{figure}


The previous section discussed how AI/ML practitioners are changing the fungibility and value of observations for weather/climate prediction. This section centers the meteorological \textit{data} sets which lie near the middle of the data pipeline, between atmospheric observations and the physics-based models that produce forecasts. Meteorological data has grown so large over the last twenty years that MIT meteorologist Kerry Emanuel recently warned of a "sidestep of theory" as researchers increasingly worked with "enormous data sets produced by remote sensing and numerical models" \cite{emanuel2020}. As AI/ML methods have emerged in meteorology, they have had to prospect and interface with these enormous meteorological datasets in ways which reveal how each field creates structured and useful data at scale. 

This section focuses on one specific problem with the initial AI/ML weather/climate models (GraphCast, PanguWeather, FourCastNet): they were all trained on the meteorological reanalysis dataset ERA-5. ERA-5 is a reanalysis dataset, a dataset of the atmosphere produced by interpreting observations using physics-based models. This meant that these initial AI/ML models were training in some sense on the output of existing meteorological models, a situation described by interviewees as "a model training a model" [P1] which limited their operational use. In attempting to solve this problem, AI/ML researchers are now attempting to detach AI/ML models from reanalysis data, to "go up the food chain" [P6] and train their models directly on observational and remote sensing datasets. This would free AI/ML models from reliance on the common data infrastructure produced by national agencies under state scale, allowing private meteorological companies to scale AI/ML models from the ground up with their own proprietary datasets. In discussing the ongoing turn from reanalysis data to satellite / remote sensing data, this section argues that AI/ML methods prioritize satellite data not because it is "ground truth" or somehow closer to observations, but because (1) it can bypass the nesting practices of meteorology's state scale and (2) it is connected to an entrepreneurial network of private satellite companies that are already expanding into AI/ML models.

\subsubsection{ERA-5 and Reanalysis Data}

Modern physics-based weather/climate models simulate the future atmosphere from the present one; to do this, the models must have a complete and physically consistent dataset of the atmosphere as it currently is and was (usually for the last few decades). This is called a reanalysis dataset, and it must be carefully generated by interpolating sparse observations into a dense spatial and temporal grid through the use of physical laws. As of 2026, all current AI/ML weather models have been trained on the European reanalysis dataset ERA-5, and this fact came up almost immediately in most [10/18] interviews once the conversation turned to AI/ML models. Per P12, a head meteorologist at a AI weather startup, one thing "that's worth stating explicitly right now, the AI modeling is not a closed loop, and it's because of this training on and reliance on ERA-5." By "closed loop," P12 means that AI/ML models still rely on meteorological analysis; the models are not capable of prediction outside of / without meteorological domain knowledge. Participants saw two main issues in this overreliance on ERA-5. The first was that AI/ML models (despite their claims of domain independence) were still dependent on meteorological reanalysis datasets in general. The second was that the AI/ML models were reliant on ERA-5 in particular: ERA-5 was a specific reanalysis dataset with quirks that might bias AI/ML model output. So the initial AI/ML models were not really independent from physics-based reanalysis, and they also had a concerning lack of robustness because of their specific reliance on ERA-5.

For many participants, the initial AI/ML models' reliance on reanalysis data meant they were not truly independent of meteorological practice, being at best an approximation of the nested world models produced by state scale. P4 said this most clearly in a back-and-forth with A1:

\begin{quote}
A1: ...but they're [the AI models] not physically bound in the same way, right?

P4: They're not, but they also kind of are [laughing]. I mean, because they're trained on climatology [long-term trends in the atmosphere]--

A1: Because they're trained on ERA-5?

P4: Exactly. They're trained on climatology. So the best that they can do is...
\end{quote}

In this exchange, P4 notes that while physical theory is not explicitly coded into AI/ML weather models, AI/ML models are trained on climatology, an embedding of meteorologists' long-term knowledge of the Earth's climate. In a sense, this meant that AI/ML models were getting the same basic physical equations meteorologists use, but mediated through meteorologists' own collective attempts to understand the atmosphere. P1, an observational meteorologist at a tech company, called this situation a 'model training a model:' "I call it a model training a model, right? I would rather have no [physical] model. I would rather have pure observations and figure this out" [P1]. P9 put it similarly: "yeah, if you're going to do AI on the model, it's going to tell you the equations that run the model" [P9]. To these interviewees, the reliance of AI/ML models on reanalysis data meant that they were still reliant on the traditional physics-based models that generated reanalysis data, at best efficient simulations of physics-based models rather than being independent models in their own right.

Simultaneously, several interviewees said that the reliance of current AI/ML methods on ERA-5 in particular was a sign of how little of the meteorological data infrastructure was "AI ready;" it represented an uncomfortable lack of robustness in the different models. P4, a CEO of a private weather company, expressed this frustration most explicitly: "Right now all these other models, GraphCast, Pangu-weather, most of them, if not all of them, are training on ERA-5. \textbf{Same damn data set}" [P4]. In noting this centralization, P8 and P12 noted that ERA-5 had artifacts and biases that were specific to ERA-5 itself ("a skew issue"), and which might be different or absent in other reanalysis datasets; this had (per P8) prompted attempts to make other reanalysis datasets AI-ready to improve robustness. P6, the CEO of an AI weather startup, identified this as a key gap in current AI/ML practice: "all these models train on the same data, and those datasets are fundamentally limited" [P6]. Put simply, current AI/ML models were not just training on weather/climate obervational data; they were training on meteorological domain knowledge via reanalysis datasets such as ERA-5. 

\subsubsection{Moving Past ERA-5 with Satellite Data}

The current push of AI/ML meteorology at the time of these interviews was to make use of observation and remote sensing datasets for AI/ML methods, bypassing reanalysis data entirely. In turn, this would allow private companies to bypass the reanalysis networks of the national meteorological agencies. In their recent paper on the subject, Allen et. al refer to this as "end-to-end" AI/ML weather prediction because bypassing reanalysis data would make the entire data pipeline AI/ML-driven. Per P12, "the whole point [of AI/ML modeling] is that we don't define the atmosphere in terms of physical components" [P12]; as we noted in the Background (2.3), AI/ML methods have distinguished themselves by deprioritizing "human knowledge of the domain," and so the current entanglement of physical theory into the data used by AI/ML models has been understood both ideologically and technically as an obstacle to their growth. In order to become "end to end" and therefore to entrepreneurially scale data directly, AI/ML models would need to become independent from reanalysis data.

This independence from reanalysis data, pursued by researchers like Allen et. al \cite{allen2025} and private weather companies like Tomorrow.io, has involved the extensive use of satellite data. Satellite data are especially convenient as input to AI/ML prediction because they are compiled directly from global-scale satellite readings. Under state scale, reanalysis data must carefully nest and assimilate sparse surface and upper-air observations via physics-based models at national and global scales. But satellite data, which is produced at national and global scales from the start, can be globally useful for a single company without having to go through the nesting practices of state scale. As P14, a head meteorologist at a private weather company, put it: "The same pixel sees the same spot every day, every 10 minutes of every day. So when people go from satellite [readings] to global forecasting, that's easy because they don't move. The geostationaries don't move" [P14]. While this becomes more complicated with polar satellites, which cut across much smaller portions of the globe, satellite readings still have an inherent scale larger than any other meteorological observation type, allowing satellite datasets to bypass the traditional way meteorological observations scale, which is via nesting smaller scales within larger ones \cite{tsing2011}.

However, while satellite readings are understood as observations by AI/ML researchers for the purpose of predicting "directly from observations," their status as observations is tentative in meteorology. Unlike direct observations of parameters like temperature, precipitation, and humidity, satellite readings only indirectly measure the ground truth state of the atmosphere, often requiring intense calibration with surface/balloon measurements. Most interviewees had something to say about satellite readings, usually placing them somewhere in between observations and models. Most bluntly, P4 said that "Satellite data is also really valuable, but it's indirect. So they're basically models." Similarly, P8 said that "Satellite observations, because they're not in situ observations, they're remotely sensed, they're literally an average of a bunch of layers of the atmosphere that we kind of turn into a profile through statistical or physical retrieval methods." P3, a meteorological researcher, noted that for several problems satellite data was used because it was the main source of data available, but simultaneously noted that satellite data was not exactly ground truth that could be verified against: "It's really hard to get the verification data... like we have satellite data, and usually that's what we use as verification. \textbf{But like, is it?}" [P3]. We use "satellite readings" as a common term for clarity, but most interviewees referred to satellite readings as observations, data, or even models in a somewhat slippery way, making it clear that satellite observations/data had a tentative connection to ground truth. 

So even though AI/ML initiatives are prioritizing satellite data, this does not seem to be because of some uncontested sense of it as ground truth; interviewees considered satellite data highly detached from ground truth, in some cases practically a model in itself. Rather, we argue that AI/ML initiatives prioritize satellite data because it aligns with AI entrepreneurial scale in two key ways. First, as discussed in Section 4.1, several of the private companies capable of making global observations (Tomorrow.io, Spire) rely heavily on satellite data because it can build global datasets without the need to own land. Because these companies also have access to capital from non-meteorological organizations like NASA and the US military, they have been able to front the compute and research requirements necessary to attempt end-to-end modeling. 

Second, despite their inability to wholly replace observations, satellite readings are designed to have a larger scale than any other observation types. Because satellites observe a dense grid of locations, satellite products do not need to undergo gridding and reanalysis in the same way as more sparse observations like balloons and ground stations; they do not need to be spatially nested like observations usually are under state scale. As such, AI/ML initiatives working with satellite data can more easily produce an end-to-end data pipeline that bypasses national meteorological agencies and their reanalysis datasets, training on satellite data that is already a bit outside meteorology's state scale.

\subsection{Scaling Models: Ensembles and Physical Consistency}
\begin{figure}
    \centering
    \includegraphics[width=1\linewidth]{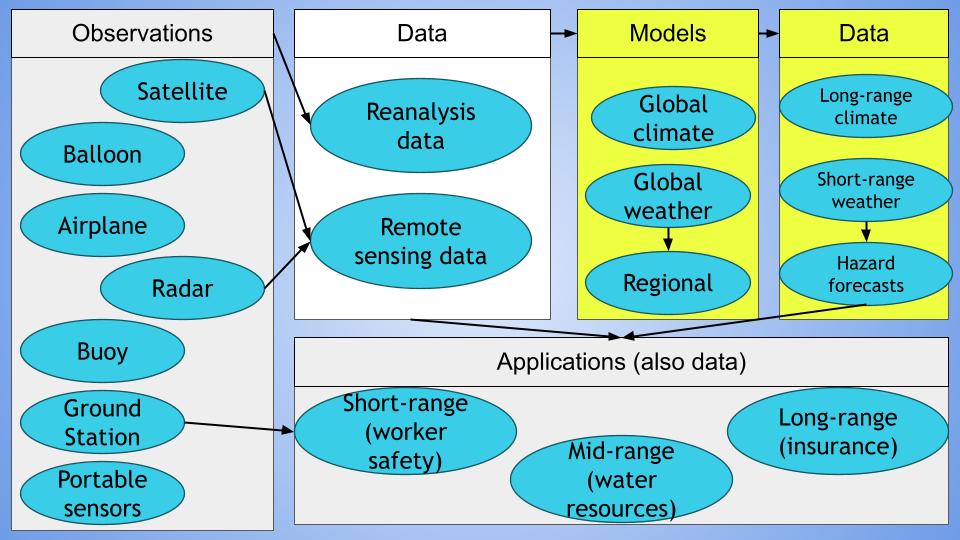}
    \caption{The current section discusses \textbf{models and model outputs.}}
    \Description[The meteorological "data pipeline," loosely split into observations, data, and models. The Models box is highlighted.]{Four boxes titled Observations, Data, Models, and Data. The Observations box contains five seven bubbles: Satellite, Balloon, Airplane, Radar, Buoy, Ground Station, and Portable Sensors. The Data box contains Reanalysis data and Remote sensing data. The Models box contains Global climate, Global weather, and Regional bubbles, stacked vertically, with each bubble pointing to the one below it. The second Data box contains Long-range climate, Short-range weather, and Hazard bubbles, representing different kinds of data products produced by models. Finally, there is an "Applications (also data)" box spanning the bottom of the figure, containing Short-range (worker safety), Mid-range (water resources), and Long-range (insurance) applications. Some observations are connected directly to the "short-range" bubble, and both "data" boxes are connected to the "Applications" box. The Models box is highlighted.}
    \label{fig:pipeline_models}
\end{figure}

The previous sections have established differences in how AI/ML and meteorology scale observations and data in their respective data pipelines. This section discusses frictions in the ways that AI/ML and physics-based meteorological models achieve scale, focusing on places where a model can be understood to scale or be scalable under AI/ML's entrepreneurial scale but not meteorology's state scale, or vice versa. First, we discuss \textit{speed and resolution}, which we understand as AI/ML scale without meteorological scale. AI/ML weather models are significantly faster than meteorological models, which implies scale in an entrepreneurial (AI/ML) sense due to an improved ability to exploit large-scale data. But this speed struggles to translate to scale in a state (meteorological) sense of representing phenomena at spatial/temporal scales. Second, we discuss how AI/ML models have attempted to work across the interdisciplinary user network of weather/climate models. Under meteorology's state scale, weather/climate models are readily understood across disciplines because of their shared reliance on physical law, allowing them to represent specific natural phenomena (forest fires, flash floods) to claim legitimacy. Because AI/ML models are detached from physical law, they have largely gained traction in industries like tech and finance which have less dependence on physical phenomena and/or rigid statistics.



\subsubsection{Speed and Resolution: AI/ML Scale without Meteorological Scale}

The most immediate affordance of AI/ML models was their ability to run exceptionally quickly, and/or at much higher spatiotemporal resolution, compared to traditional GCMs. For example, P6, a meteorological startup founder, argued that "What became pretty clear was that, okay, we can approximate the accuracy of the numerical models for 10s of 1000s of times less cost, and that's going to create business opportunity that people don't yet really understand, because that kind of pattern has happened several times with different technologies in the past" [P6]. All three of the initial AI/ML forecasting papers explicitly highlight speed and/or resolution as a key benefit of AI/ML models: Google's Graphcast "predicts hundreds of weather variables, over 10 days at 0.25° resolution globally, in under one minute" \cite{lam2023}, Huawei Cloud's PanguWeather was "more than 10,000-times faster than the operational IFS [physics models]" \cite{bi2023}, and Nvidia's FourCastNet was "orders of magnitude faster than IFS [physics models]" \cite{pathak2022}. Under AI/ML entrepreneurial scale, where scalability is understood (per the GraphCast paper) as the ability for a model to improve with "greater compute resources and higher resolution data" \cite{lam2023}, speed and attachment to GPU infrastructures directly enables AI/ML models to scale in an entrepreneurial sense. Across A1's interviews, however, we tended not to see any particular association between this resolution/speed improvement and meteorology's state scale. 

Interviewees noted many valuable benefits of faster models. For forecasting, P1 argued that AI/ML models could immediately pull data through to forecasts: "obviously, NWP [physics-based modeling] folks have been doing this for years. And they see AI as a challenge. Now, AI models are not the greatest, they're still on the road to get better. But as a forecaster, if I had that data in front of me right away, I'll be like, yeah! I can make better decisions faster. I don't have to wait. I don't have to wait." The speed of AI models meant that rather than post-processing a model with new observations via the use of statistical hacks, new observations could quickly be pulled directly into the core model. P11, an atmospheric scientist working on air pollution, argued that this speed could make AI/ML models particularly useful for time-sensitive public-facing forecasts: "Doing 48 to 70 hour air quality forecasts to support warnings and alerts, basically, how do we provide faster information to communities? AI/ML can probably help with that" [P11]. Other interviewees noted a wide range of other benefits, such as massive ensembles [P10] and location/domain-specific finetuning [P12]. But none of this seemed to link explicitly to meteorological scale: they were questions of efficiency, resolution, and forecasting density which seemed independent of how models connected to the spatial ("global-scale," "meso-scale," "synoptic-scale") or temporal ("daily, weekly, seasonal") scales used in meteorology. To P12, it was crucial to move past "a value proposition of, oh, the AI models are faster," which (as mentioned in Section 5.2) is closely tied to an understanding of AI/ML models as just efficiently simulating physics-based models. As AI/ML weather models attempted to mature, it became crucial to take AI/ML models' traits that were scalable in an entrepreneurial sense (speed, ability to utilize optimized GPU infrastructures) and make them imply scale in the sense of meteorology's state scale.

Fortunately, meteorology is an established Big Data system with operations (like ensembling) that trade off the efficiency of a weather model for predictability or applicability at scale. For example, ensembles (the combination of many models into a more robust set of predictions) have traditionally been used to trade model speed for model predictability by showing how a model is sensitive to its initial conditions, enabling a host of negotiations around "how confident or uncertain your model is in making the forecast" [P3]. On paper, AI/ML models could use ensembles to tradeoff their speed into predictability and therefore become usable to predict large-scale phenomena such as droughts or long-term climate. Unfortunately, AI/ML models have struggled to use these established routes due to their different internal structure. Because ensembles of AI/ML models lack physical consistency, per P12, "when people say oh, it's so fast, we can run thousands of simulations [ensembles] in the same time that we would run one [physical model]. Well, a reasonable question to ask is, what's the value of doing that? What actual posterior distribution are you sampling from?" The function of ensembles in physics-based meteorology does not seem to automatically carry over to AI/ML-based meteorology. Physics-based models destabilize in a particular way over time, "blowing up" as error propagates through their physical equations, but AI/ML models change over time in much less predictable ways. P4,P8, and P12 all noted different and bizarre ways that AI/ML models evolved; P8 described strange stationary waves appearing across parameters like humidity and pressure which had no clear physical root but seemed to be caused by the convolutional structure of the AI/ML model itself. These unpredictable ways that models failed meant that despite AI/ML model ensembles being immediately promising to many interviewees [P4,P10,P11,P12], they did not automatically have deep value in the same sense as ensembles of meteorological models. 

\subsubsection{Interdisciplinary Networks of Use in Meteorology}

Simultaneously, efforts to make AI/ML models useful have revealed the critical role of physical consistency in how meteorological models are used operationally across domains like forestry, oceanography, and water resource management. P4 described an AI forecasting workshop attended by operational forecasters and a GraphCast researcher (from Google Deepmind); operational forecasters variously said, "I can't use this. There's no physical consistency." The AI/ML researcher replied, "You tell me the metrics you're interested in, and I'll optimize for those metrics." Here we see a classic mismatch between AI/ML and meteorological senses of model usefulness/trust. While AI/ML research has a healthy distrust of benchmarks in its own right \cite{cooper2023,henderson2019,liao2022}, the solution is typically more or better benchmarks. In contrast, meteorological models are often trusted across disciplines based on a qualitative sense of "model skill," a model's ability to properly represent phenomena at different scales. Additionally, downstream models (such as air quality, hydrological, and oceanographic models) often assume the physical consistency of the input data. Without that consistency, "do we make all of those models AI too?!" [P4]. AI/ML models struggled to travel across disciplines in the ways typical to physics-based meteorological models because physical consistency had operated as a kind of common assumption across many different fields. In this context, rather than AI/ML models augmenting or replacing a single isolated domain, AI/ML methods appeared to P4 as a necessarily contagious force that would have to replace not just meteorological models but the models of all of meteorology's downstream users such as finance, actuarial science, water resource management, ecology, and more, replacing the entire interdisciplinary data network around meteorology.

This sense of contagion may help explain why initial AI/ML weather models have integrated most readily into tech platforms (Google Search, Microsoft Aurora) and into financial models (given that since roughly 2016 machine learning has been integral to financial practice, \cite{hansen2023}), while they have largely failed to penetrate more classically statistical markets like reinsurance. To P14, an AI/ML researcher and head meteorologist at a major climate reinsurance company, trust and reliability demands often put AI/ML modelers at odds with more established actuarials ("statistics nerds"): "The head of data science at my company, he and I argue about this all the time. He's like, my ideal model is a model that has no predictors [is as simple as possible] and gives you a perfect forecast. But I've already accepted the -- Doctor Strangelove, right? How I learned to stop worrying and just love the atomic bomb?" [P14] In describing the difference between himself and more traditional statisticians, P14 refers to a classic satire film where the eponymous Doctor Strangelove rationalizes and makes peace with impending nuclear war. In the context of AI/ML, this means accepting the opacity and uncertain outputs of AI/ML models, giving up on the traditional statistician's search for simple and explainable models. Compared to both meteorology and conventional statistics, the use of AI/ML models is understood to demand specific orientations towards uncertain model outputs and therefore error, a new tolerance for what were previously signs of fundamental model failure (such as outputs like "where it's 80 degrees and two feet of snow" [P4]).

So AI/ML model forecasts were unable to move across disciplines in the same way as the forecasts of physics-based models, both because adherence to physical law is a kind of interdisciplinary common ground, and because the data science standards of fields like reinsurance clash with the opacity and uncertainty inherent to AI/ML methods. The result of this was that, as P10 noted, the current AI/ML models have been used mostly "for consumers, I think they've really focused on." AI/ML models have been served through tech platforms at Google and Microsoft, and they have been used by tech startups aiming to disrupt existing industries, but they have largely failed to utilize existing systems of interdisciplinary travel (both public and private) in meteorology. Rather than just being a kind of institutional conservatism, this section argued that this is because of a mismatch between what makes models usable at scale between AI/ML (entrepreneurial scale) and meteorology (state scale). 

\section{Discussion: What does the "Regimes of Scale" framework mean for HCI?}

Our paper's framework of regimes of scale shows that despite their domain-agnostic framing, modern AI/ML methods operate in practice as a deeply situated way of knowing and theorizing about the world. AI/ML developed as a regime of scale in order to exploit the influx of platform and internet data that appeared over the 2000s and 2010s, and as it has come into contact with meteorology, which developed in response to global observation networks over the 1960s-80s, the two fields' different ways of scaling observations/data/models have come into conflict as AI/ML weather models attempt to prospect observations [4.1], to bypass conventional data organizing strategies (reanalysis) [4.2], and to move using meteorology's own interdisciplinary networks [4.3]. 

This Discussion section makes two contributions to HCI research on AI/ML applications. First, rather than understanding AI/ML as taking over, collaborating, or extracting from domain sciences, we advance a relational analysis of AI/ML application to domains. Understanding AI/ML in relation to existing modes of data organizing allows us to identify how AI/ML methods break down when encountering other regimes of scale and also demystify the notion that AI/ML is domain-agnostic. Second, rather than understanding AI/ML as being applied to an isolated domain, we show how AI/ML methods have been forced to interface with and rework the entire network of users around meteorology. As a regime of scale, meteorology has existing ways to apply its models and forecasts across domains, and AI/ML methods have appropriated and renegotiated these networks while bringing in entirely new networks of users.


\subsection{The Importance of Relational Analysis for AI Applications: AI/ML in Friction with Rival Big Datas}

Ever since Star's classic work on the Berkeley Museum of Vertebrate Zoology, which established how a set of methods and a collection of boundary objects could enable communication between different social worlds \cite{star1989}, HCI has advanced research on how standards and information systems enable interdisciplinary collaboration without eliminating important differences within each field. This line of research has expanded on traditional laboratory studies by studying large-scale infrastructures or information systems, such as the Worm Community System \cite{star1994} or the WATer and Environmental Research Systems (WATERS) cyberinfrastructure project \cite{ribes2010}, as they aim to function across many different disciplines or institutions (see also \cite{hartswood2012,neang2023,ribes2026}). Our central contribution in this paper is to put two of these interdisciplinary collaboration systems (which we call "regimes of scale") side by side and analyze them relationally as their standards, senses of good data and bad data, and institutional networks come into contact with one another. This allows us to conduct a relational analysis of scientific infrastructure similar to Andrew Abbott's relational analysis of professions \cite{abbott1988}. Abbott argued that professions are best understood by studying jurisdictional disputes, the places where multiple professions vie for legitimacy over a single problem. In the same way, we focus not just on the oft-cited rhetoric of data science / AI/ML infrastructure overtaking or imposing on the domain sciences but on the places (as in meteorology) where they must contest fundamentally different ways of organizing data at scale.

The privatization of basic AI/ML research \cite{ahmed2023} and data/model infrastructures \cite{widder2023}, as well as a large influx of startup capital to "disrupt" existing industries, makes our relational analysis via regimes of scale particularly relevant to current HCI work on AI/ML and data-scientific applications. Contemporary AI/ML and data science application research has focused on the contrast between situated domains and domain-agnostic data science \cite{slota2020}. In this literature, domains are essentially the local to data science / AI/ML's global, which is why the representation of domain knowledge can be understood as an effective method to improve actionability of AI/ML models \cite{jung2024} or to get timely feedback on deployments of AI/ML \cite{sambasivan2022}. We bypass the framework of domain-agnostic versus domain science in favor of regimes of scale because it allows us to highlight how AI/ML integration into meteorology has been shaped according to meteorology's existing regime of scale. This is not to say that AI/ML has had to conform to meteorological ways of organizing data or the meteorological data pipeline: for example, our Findings [4.1] show how ground observations devalued in meteorology have been regarded as potential key assets for AI/ML models. But it is precisely through a relational analysis that we are able to identify how AI/ML weather models negotiate and reconfigure the value of different data, observations, and models in ways that may at times align nicely with the interests of meteorologists (personalized forecasts, fast models in section 4.3) while also doing things that are seemingly contradictory under the state regime of scale (using ground observations for large scale modeling in Section 4.1). The contrasts between the scale-making practices of entrepreneurial and state scale, where things scale under AI/ML but not under meteorology and vice versa, are the places that shed light on both AI/ML and meteorology's specific ways of organizing data infrastructures across disciplines and institutions.

This paper's relational analysis of the AI/ML and meteorological big data regimes was possible because both AI/ML and physics-based meteorological models are attempting to solve nearly identical weather/climate problems. Rather than studying one information infrastructure across many domain applications, this paper has studied multiple information infrastructures across one (or a small set of) common applications. Other regimes of scale, such as hospital data systems, biometric systems like Aadhaar, and the scientific infrastructure projects HCI has studied over the 2000s and 2010s are now potentially interfacing with the AI/ML regime of scale. Relational analysis via regimes of scale can help HCI understand what happens when AI/ML methods come into contact with these preexisting ways of scaling data, and can help identify these situated practices of scaling and theorizing about the world. 

\subsection{AI/ML in Friction with Other Networks of Users}

The previous section discussed the benefits of relational analysis (via regimes of scale) for demystifying AI/ML's supposed disruption of the domain sciences. This section discusses the second contribution of regimes of scale: their ability to account for AI/ML and meteorology's different networks of users, even as these methods are used for very similar problems. Even before the 2022-2023 AI/ML weather models, tech companies had already produced weather/climate products for interdisciplinary use, most notably the environmental and geospatial data analysis platform Google Earth Engine. But as P12 noted, this seemed to be speaking to a completely different set of users than the user network in meteorology: "There is a non expert group that might prefer going to it, but I have actually never met anybody who uses any of the first party platform copies of the data [Google Earth Engine]. Every single person I've ever met, even non meteorologists, they still go directly to the NOAA web servers to access this type of data." In the same way, the new AI/ML models represented not just a different way of organizing data but a different set of users with different expectations for good/bad data. Across observations, data, and models, we see how AI/ML represented a different way to make components of the data pipeline usable to finance companies, tech companies, and AI/ML practitioners themselves, displacing existing user networks for meteorological data.

Most notably, Section 4.1 showed how AI/ML has been heavily associated with lending an intrinsic value ("data is the new gold") to weather/climate observations and enabling their sale to private users. This involves private weather observation companies, who previously mainly sold weather observations to NOAA, utilizing AI/ML to directly convert their proprietary observations into valuable forecasts. Simultaneously, AI/ML practitioners appeared externally as potential buyers of weather observation data (see P5's quote), even if there was no clear line to profit outside of NOAA. This is in line with Burcu Baykurt's analysis of Gov-tech as capture, which understands government tech as a means of turning fragmented data into packaged data products that are newly valuable for the market \cite{baykurt2025}. In this sense, AI/ML methods can be understood as mining (or perhaps prospecting) state observation infrastructures, building on the 1990s partial privatization of weather/climate, and opening them up to new sets of buyers.

However, as Section 4.2 notes, this marketization of observations can come at the cost of the user base for NOAA's data products and forecasts. Any observations packaged and sold on the open market cannot directly be part of NOAA's publicly-available data infrastructure, which has historically defined the interdisciplinary use of meteorological data in the US. Fields like insurance and finance rely on a common base of publicly-available weather observations and model forecasts, often due to liability concerns; for example, P14 noted that most reinsurance companies kept proprietary "data about people who are insured" while using common weather products like GFS. As observations become part of a value chain for private companies, they are also removed from centralized institutions that govern their production and use in order to produce common data infrastructures. Future work in HCI should consider how increased data fungibility and new networks of users for data can lead to new forms of data misuse \cite{pasquetto2024} and can cause a kind of balkanization of centralized data infrastructures as institutions like NOAA are less able to cheaply expropriate data for aggregation into a common data infrastructure.

\section{Conclusion}
Through interviews with 18 forecasters/meteorologists from a range of public and private institutions, this paper examines how two different regimes of scale -- the meteorological state scale and the AI/ML entrepreneurial scale -- produce frictions across meteorological observations, data, and models. We argue that despite the nominally domain-agnostic nature of AI/ML methods, they operate in terms of an entrepreneurial scale or a recursive user-data-model relationship that scales a tech company's available data and the power of its proprietary models as it collects more data from its human users. This ties AI/ML methods to specific arbitrage/speculation applications such as advertising, finance, and platform administration. But this entrepreneurial regime of scale faces frictions when encountering methods that meteorology has used to scale their own observations, data, and models, such as physical consistency and the observation and model infrastructures of national agencies. This paper's central argument is that meteorology, and many of the so-called domains of AI/ML application, have their own existing ways of scaling Big Data. As AI/ML is applied to the domains, these preexisting ways of organizing data shape and are shaped by AI/ML's entrepreneurial regime of scale, producing many of the frictions we see in AI/ML in practice.


\bibliographystyle{ACM-Reference-Format}
\bibliography{sample-base}


\end{document}